\documentclass{article} 
\usepackage{nips13submit_e,times}
\usepackage{changepage}
\usepackage[utf8]{inputenc}
\usepackage{graphicx}
\usepackage{amssymb,amsmath,mathbbol}
\usepackage[margin=2.5cm]{geometry}
\usepackage{rotating}
\usepackage{textcomp,marvosym}
\usepackage{multirow}
\usepackage{fixltx2e}
\usepackage{cite}
\usepackage{color}
\usepackage{microtype}
\DisableLigatures[f]{encoding = *, family = * }
\usepackage[hyphens]{url}
\usepackage[hidelinks]{hyperref}
\hypersetup{breaklinks=true}
\newcommand{\ignore}[1]{}

\definecolor{myblue}{rgb}{0.2,0,0.5} 
\definecolor{mypurple}{rgb}{1.0,0,0.4} 

\title{Seasonal Linear Predictivity\\ in National Football Championships}

\author{G. Jurman\\ 
Fondazione Bruno Kessler, Trento, Italy \\
jurman@fbk.eu
}
\date{}

\nipsfinalcopy 

\begin{document}
\maketitle

\bibliographystyle{plos2015}

\makeatletter
\renewcommand{\@biblabel}[1]{\quad#1.}
\makeatother

\begin{abstract}
Predicting the results of sport matches and competitions is an arising research field, benefiting from the growing amount of available data and the novel data analytics techniques.
Excellent forecasts can be achieved by advanced machine learning methods applied to detailed historical data, especially in very popular sports such as football (soccer).
Here we show that, despite the large number of confounding factors, the results of a football team in longer competitions (\textit{e.g.}, a national league) follow a basically linear trend useful for predictive purposes, too.
In support of this claim, we present a set of experiments of linear regression on a database collecting the yearly results of 707 teams playing in 22 divisions from 11 countries, in 20 football seasons.
\end{abstract}

\section*{Introduction}
Predicting sport results in the last few years has ceased being only almost an art for initiated specialists~\cite{gonzalez10predicting} to enter the realm of data analytics, thus providing a further support to the claim of considering as science many aspects of several sports~\cite{wesson02science,dobson11economics}.

In particular, interest in forecasting sport competitions' results has grown in the last few years essentially because of two key factors: the arising need for more realiable predictive models by the betting agencies~\cite{langseth13beating,goddard04forecasting,sheridan12modelling,worton14predicting,crowder02dynamic}, and the increasing number of available sources collecting data at different level of details.
However, the predictability of the results is still a debated issue~\cite{buursma11predicting,mchale10science,heuer10soccer,zyga10can,mit09statistical}, mainly because of the random effects affecting the outcome of a match, with football (soccer) as a major example~\cite{peel92demand,colwell82random,skinner09soccer,eastwood14how,duin14are,altman13we}.
Many algorithms from statistics and machine learning have been recently used to overcome such randomness bias so to achieve good predictive performance~\cite{cattelan13dynamic,haghighat13review,dobson03persistence,baio10bayesian,owramipur13football,min08compound,constantinou13determining,robinson10simple,heuer12how,heuer14optimizing}, either applied to data catching diverse aspects of the game, or with different historical span or at various level of details.
For instance, novel approaches are focussing on the performance of each player~\cite{linde14predicting}, or considering the complex network of all ball passes during a match~\cite{clemente15using,grund12network}.
In general, when powerful learning methods and/or a substantial wealth of training data are used, the predictive accuracy that can be reached is excellent, and the occurring randomness is effectively dealt with, even using recent social network interactions~\cite{kampakis14using,baath13modeling}.

In this paper we want to demonstrate that, despite the existing randomness and other confounding factors, there are situations where the sport results are driven by very simple (for instance, linear) trends, and these trends can be captured by basic techniques and limited amount of training data.
As in~\cite{rue00prediction} we focus on a longer competition such as a national league, and we show the outcome of forecasting the last part of a season by using only the results of the initial portion of the campaign.
Here we restrict to national football (soccer) championships and the simplest possible (predictive) statistical techinque, \textit{i.e.}, linear regression as in~\cite{goddard05regression,figueiredo14how}.
Note that linear regression has already been used to forecast future league points, using as predictors some economical indicators such as turnover, profit/loss before tax, net debt, interest owed on any debt and the club's wage bill~\cite{ellis13how}.
In particular, we want to assess to which extent such a simple approach used only on the current season results, without any historical data, can be effectively used to predict the behaviour of a team in the final portion of a tournament, both in terms of the total number of earned points and the final ranking in the championship table.


\section*{Analysis}

\subsection*{Data description}
\begin{figure}[!t]
\begin{center}
\includegraphics[width=0.8\textwidth]{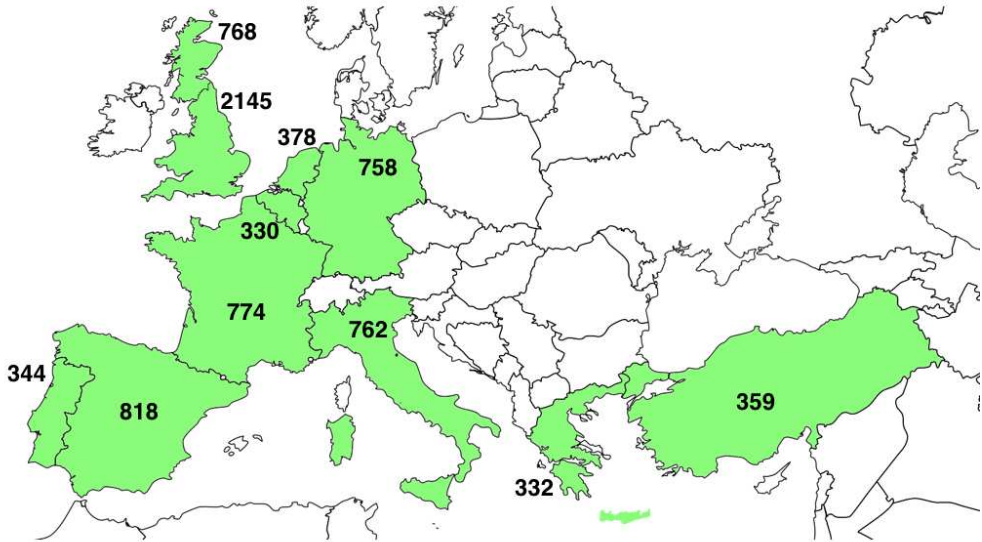}
\caption{Geographical distribution of the 7768 time series in the database.}
\label{fig:map}
\end{center}
\end{figure}

Data are extracted from the Football-Data repository~\cite{football15data} and they include the results of all matches for 425 european national championships, over the 21-years time range 1993/94--2013/14.
In detail, data for 22 divisions at different levels are studied, for a total of 7768 series for 707 unique teams: championships grouped by league and number of matchdays are enumerated in Tab.~\ref{tab:rounds}, while distribution of the 7768 time series by country is shown in Fig.~\ref{fig:map}.

\begin{table}[!ht]
\begin{center}
\caption{Tournaments grouped by number of rounds and league}
\begin{tabular}{ll|rrrrrrrrrr|r}
\multicolumn{2}{c|}{Championship} & \multicolumn{11}{c}{Rounds}\\
Country& League &  26&  28&  30&  32&  34&  36&  38&  42&  44&  46&  Total \\
\hline
Belgium & Pro League&   &  1&  4&  1&  13&   &   &   &   &   & 19\\
Germany & Bundesliga&   &   &   &   &  21&   &   &   &   &   & 21\\
Germany & Zweite Bundesliga&   &   &   &   &  2 &   &  1&   &   &   & 21\\
England & Premier League&   &   &   &   &   &   &  19&  2&   &   & 21\\
England & Championship&   &   &   &   &   &   &   &   &   &  21 & 21\\
England & League One&   &   &   &   &   &   &   &   &   &  21 & 21\\
England & League Two&   &   &   &   &   &   &   &  2&   &  19 & 21\\
England & National League&   &   &   &   &   &   &   &  1&  1&  7 & 9\\
France &Ligue 1&   &   &   &   &  5&   &  16&   &   &   & 21\\
France &Ligue 2&   &   &   &   &   &   &  16&  2&   &   & 18\\
Greece & Superleague&  1&   &  12&   &  7&   &   &   &   &   & 20\\
Italy & Serie A&   &   &   &   &  11&   &  1 &   &   &   & 21\\
Italy &Serie B&   &   &   &   &   &   &  6&  1 &   &  1 & 17\\
The Netherlands & Eredivisie&   &   &   &   &  21&   &   &   &   &   & 21\\
Portugal & Primeira Liga&   &   &  8&   &  12&   &   &   &   &   & 20\\
Scotland & Premiership&   &   &   &   &   &  6&  14&   &   &   & 20\\
Scotland & Championship&   &   &   &   &   &  2 &   &   &   &   & 20\\
Scotland & League One&   &   &   &   &   &  17&   &   &   &   & 17\\
Scotland & League Two&   &   &   &   &   &  17&   &   &   &   & 17\\
Spain & Liga&   &   &   &   &   &   &  19&  2&   &   & 21 \\
Spain & Liga Adelante&   &   &   &   &   &   &  1&  17&   &   & 18\\
Turkey & S{\"u}per Lig&   &   &   &  1&  19&   &   &   &   &   & 20\\
\hline && 1&   1 & 24 &  2 &  129&  60 & 102&  36&   1&  69 & 425 
\end{tabular}
\label{tab:rounds}
\end{center}
\end{table}

\begin{figure}[!b]
\begin{center}
\includegraphics[width=0.8\textwidth]{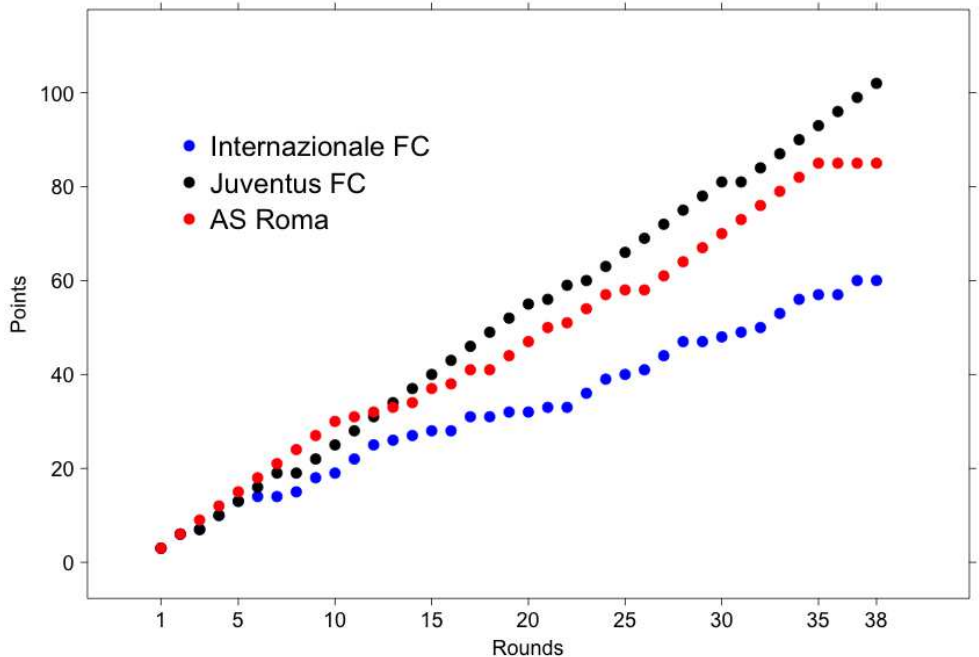}
\caption{{Time series of the points earned by Juventus FC (black), AS Roma (red) and Internazionale FC (blue) during the 2014/15 Serie A campaign}. On the $x$-axis the 38 matchdays and on the $y$-axis the accumulated points.}
\label{fig:ts_ex}
\end{center}
\end{figure}

For our purposes, all the 7768 time series are described by the two variables \textit{rounds} and \textit{points}, keeping track of the accumulated points gained by a team during the rounds of a season-long campaign, as shown in Fig.~\ref{fig:ts_ex}.

\subsection*{Methods}
All linear and polynomial predictive models are computed by the \textit{lm} function of the \textit{stats} package in the R environment~\cite{R2015}, as \textit{points} versus a linear/polynomial expression of \textit{rounds}.

Confidence intervals are computed via the Student's bootstrap procedure~\cite{efron87better,diciccio96bootstrap}, in the version described in~\cite{davidson97bootstrap} and implemented in the \textit{boot.ci} function of the \textit{boot} R package. 

\begin{figure}[!t]
\begin{center}
\includegraphics[width=0.95\textwidth]{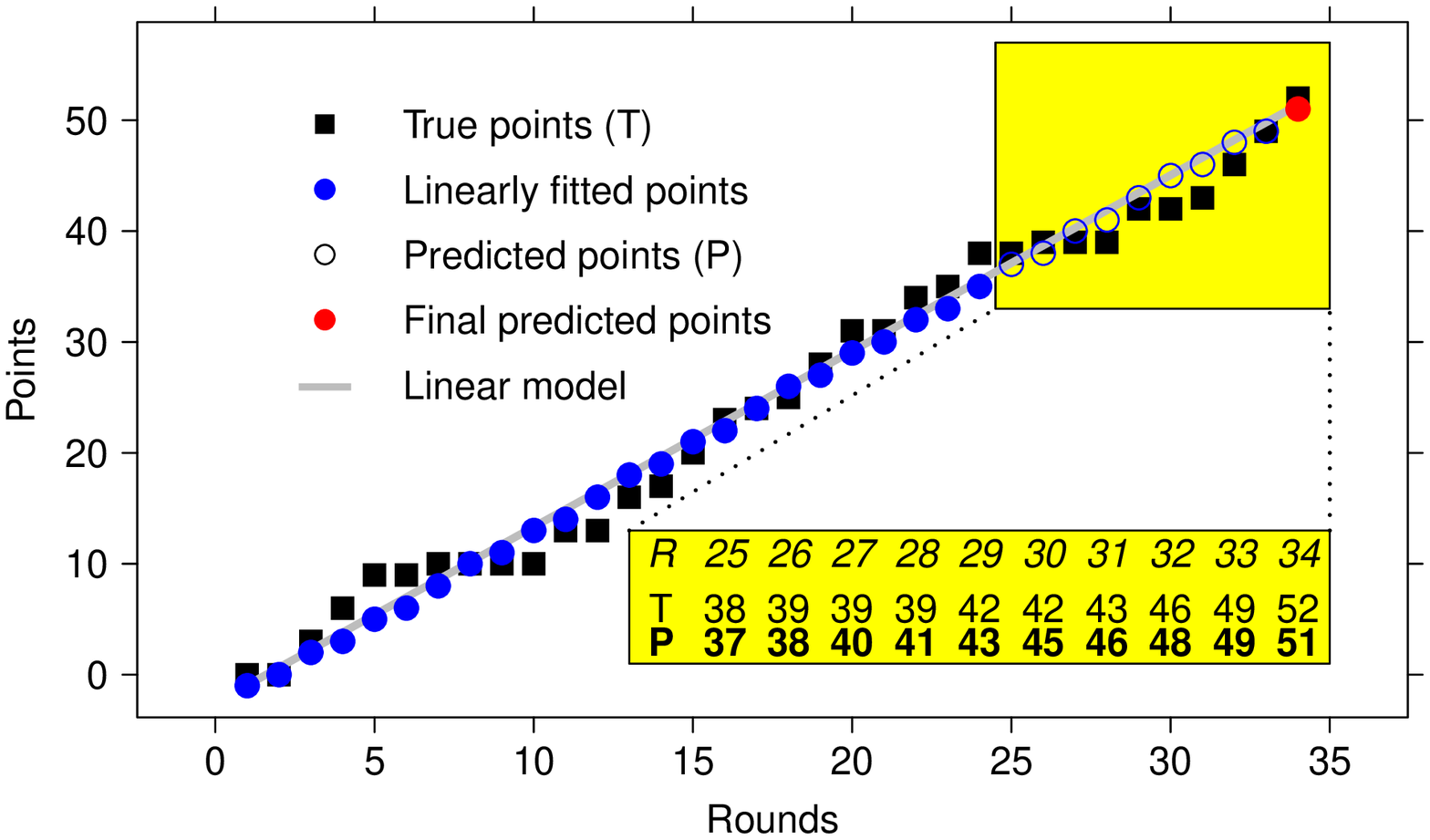}
\caption{{Points earned by Schalke 04 in the Bundesliga 2013/14 season (T, black square) and their approximation (circles) through a linear model (grey line)} trained on the first 24 rounds (blue filled circles) and extrapolated on the last 10 rounds (P, white and red circles), highlighted in the yellow box. In the bottom right yellow table, the comparison between the real points (T) and the predicted points (P) on the last 10 rounds.}
\label{fig:schalke}
\end{center}
\end{figure}

In detail, let $T$ be a team partecipating in a league whose season consists of $n$ rounds, and let $T_i$ be the number of points earned by $T$ after the $i$-th round, so that $T_n$ is the total number of points at the end of season.
Let $t_s$ an integer between 1 and $n-1$, and let $L_T^{t_s}$ be a model trained on $(1,T_1),\ldots,(n-t_s,T_{n-t_s})$. 
Define then $\bar{T}_n = \lfloor L_T^{t_s}(n) \rfloor$ as the estimated number of total points earned by $T$ as the largest integer smaller than the extrapolation of $L_T^{t_s}$ computed on the point $n$.
In Fig.~\ref{fig:schalke} an example is shown for the linear modeling of Schalke 04 season in the Bundesliga 2013/14, where the final number of earned points is predicted for $t_s=10$.

Finally, quantitative comparison between tournament standings (predicted and actual) is computed by mean of total absolute displacement of the corresponding rankings. 
Let $\mathcal{T}=\{z_1,\ldots,z_n\}$ be the teams involved in a given tournament. 
Consider now the standing $S$ after a certain matchday of the tournament, that is, the ranked list $S=[z_{\alpha_1},\ldots.z_{\alpha_n}]$ for $\{\alpha_1,\ldots,\alpha_n\}=\{1,\ldots,n\}$. 
Let $\textrm{rk}_S$ be the ranking map, \textit{i.e.} the function associating to each team $z_i$ its position inside the standing $S$, and define $\tau_S=(\textrm{rk}_S(z_1),\textrm{rk}_S(z_2),\ldots,\textrm{rk}_S(z_n))$. 
Then $\tau_S$ is a permutation of the first $n$ natural numbers, \textit{i.e.}, a member of the symmetric group $\mathcal{S}_n$; thus, to each one of all possible $n!$ standings S is biunivocally associated a permutation $\tau_S$.
Given two standings $R,S$ on $\mathcal{T}$, we define the distance $D$ between $R$ and $S$ as the total absolute displacement between $\tau_R$ and $\tau_S$:
\begin{displaymath}
D(R,S) = \sum_{i=1}^n |\textrm{rk}_R(z_i)-\textrm{rk}_S(z_i)|=\sum_{i=1}^n  |\tau_R(i)-\tau_S(i)|\ .
\end{displaymath}
In order to meaningfully compare distances computed in tournaments with different number of competing teams, $D$ is normalized by its maximum value, as computed in~\cite{mitchell04maximal}
\begin{displaymath}
\begin{split}
\max_{\tau_S,\tau_R\in\mathcal{S}_n} D(R,S) &= \max_{\tau_R\in\mathcal{S}_n} D(\textrm{Id},R)\\
&= \max_{\tau_R\in\mathcal{S}_n} \sum_{i=1}^n  |i-\tau_R(i)| \\
& = \left\lfloor\frac{n^2}{2}\right\rfloor\ , 
\end{split}
\end{displaymath}
where $\textrm{Id}$ is the identical permutation. 
We can thus define the normalized distance $d$ as follows:
\begin{displaymath}
d(R,S) = \frac{D(R,S)}{\displaystyle{\max_{\tau_S,\tau_R\in\mathcal{S}_n} D(R,S)}} 
= \frac{D(R,S)}{\displaystyle{\left\lfloor\frac{n^2}{2}\right\rfloor}} 
= \displaystyle{\frac{\displaystyle{\sum_{i=1}^n |\tau_R(i)-\tau_S(i)|}}{\displaystyle{\left\lfloor\frac{n^2}{2}\right\rfloor}}} \ .
\end{displaymath}
Furthermore, computing the expected value of $d$ over the whole permutation group $S_n$ allows the comparison of a given value of the normalized distance with the null hypothesis of the distance with a random standing:
\begin{displaymath}
\begin{split}
\mathbb{E}_{\mathcal{S}_n}(d) &= \frac{1}{|\mathcal{S}_n|} \sum_{\tau\in\mathcal{S}_n} d(\textrm{Id},\tau) \\
&= \frac{1}{n!}\frac{1}{\lfloor\frac{n^2}{2}\rfloor}  \sum_{\tau\in\mathcal{S}_n} \sum_{i=1}^n |i-\tau(i)| \\
&= \frac{1}{n!}\frac{1}{\lfloor\frac{n^2}{2}\rfloor}  \sum_{i=1}^n \sum_{\tau\in\mathcal{S}_n} |i-\tau(i)| \\
&= \frac{1}{n!}\frac{1}{\lfloor\frac{n^2}{2}\rfloor}  2 \sum_{i=1}^n \sum_{j=0}^n (n-1)! j \\
&= 2\frac{(n-1)!}{n!}\frac{1}{\lfloor\frac{n^2}{2}\rfloor} \sum_{i=1}^n \frac{(i-n-1)(i-n)}{2} \\
&= \frac{1}{n} \frac{1}{\lfloor\frac{n^2}{2}\rfloor} \frac{(n-1)n(n+1)}{3} \\
&= \frac{n^2-1}{3{\lfloor\frac{n^2}{2}\rfloor}}\\
&= \frac{2}{3}-\frac{2}{3n^2}\cdot (n\mod 2)\ ,
\end{split}
\end{displaymath}
which is $\frac{2}{3}$ for odd $n$'s and $\frac{2}{3}-\varepsilon_n$ for even $n$'s, with $\varepsilon_n$ positive, decreasing to 0 and smaller than $0.0\bar{6}$ for $n\geq 10$.
Thus, regardless of the number of playing teams, the distance $d$ between two standings in the same championship is a number ranging between $0$ (for identical rankings) and $1$ (for maximally different standings), with $d\approx \frac{2}{3}$ for randomly chosen standings.
Hereafter we show an example of the use and the interpretation of the distance $d$.

\subsection*{Example} Suppose we want to assess the error of a predictive algorithm $\mathcal{P}$ in forecasting the standing of a tournament after a given matchday, using metric $d$ as the evaluation measure.
In particular, we test $\mathcal{P}$ in two situations: (a) round 20 of italian Serie A 2014/15 and (b) the final round (18) of the South American qualifiers for the 2010 FIFA World Cup.

\paragraph{(a)} Italian Serie A 2014/15 involved 20 teams, composing the set $\mathcal{T}$ as shown in Tab.~\ref{tab:seriea}. The initial assigment of the $z_i$ labels with the team is arbitrary, and any other choice would work instead.

\begin{table}[!b]
\begin{center}
\caption{{The set $\mathcal{T}$ of teams playing in italian Serie A 2014/2015, alphabetically sorted.}}
\begin{tabular}{r|c||r|c}
Index & Team name & Index & Team name \\
\hline
$z_{1}$  & Atalanta & $z_{11}$ & Lazio \\  
$z_{2}$  & Cagliari & $z_{12}$ & Milan \\   
$z_{3}$  & Cesena & $z_{13}$ & Napoli \\   
$z_{4}$  & Chievo & $z_{14}$ & Palermo \\   
$z_{5}$  & Empoli & $z_{15}$ & Parma \\   
$z_{6}$  & Fiorentina & $z_{16}$ & Roma \\   
$z_{7}$  & Genoa & $z_{17}$ & Sampdoria \\   
$z_{8}$  & Hellas & $z_{18}$ & Sassuolo \\   
$z_{9}$  & Inter  & $z_{19}$ & Torino \\   
$z_{10}$ & Juventus & $z_{20}$ & Udinese \\  
\end{tabular}
\label{tab:seriea}
\end{center}
\end{table}

After round 20, the table, labeled as A, read as reported in Tab.~\ref{tab:rankings}. 
Suppose now that algorithm $\mathcal{P}$ predicts the championship table as in Tab.~\ref{tab:rankings}, labeled as P.
First step in evaluating the difference between standings A and P is the derivation of the corresponding permutations $\tau_A$ and $\tau_P$, and then the computation of the sum of all displacements $\tau_A-\tau_P$: as shown in the last row of Tab.~\ref{tab:rankings}, this reads as 
\begin{displaymath}
D(A,P) = \sum_{i=1}^{20} |\tau_A(i)-\tau_P(i)|=38\ ,
\end{displaymath}
thus the final normalization provides the value of the distance $d$:
\begin{displaymath}
d(A,P) = D(A,P) \cdot \frac{1}{\lfloor\frac{n^2}{2}\rfloor} = 38 \cdot \frac{1}{\frac{20^2}{2}} = \frac{38}{200} = 0.19\ ,
\end{displaymath}
which is a small number, indicating a good similarity between standings A and P, quite distant from the random value $0.\bar{6}$.

\begin{table}[!t]
\begin{center}
\caption{{Actual (A) and predicted (P) table of Serie A 2014/15 after matchday 20, with the corresponding permutations $\tau_A$ and $\tau_P$ computed with respect to the set of teams $\mathcal{T}$.} In the last column the absolute displacement $|\tau_A-\tau_P|$ is reported between A and P for the corresponding team $z_i$, and its total is indicated in the last row.}
\begin{tabular}{r|c|cp{1cm}r|c|r|r|r}
Pos. & A & P & & T & Team & $\tau_A$ & $\tau_P$ & $|\tau_A-\tau_P|$ \\
\cline{1-3}\cline{5-9}
 1 & Juventus  & Juventus  & & $z_{1}$  & Atalanta  & 15 & 14 & 1 \\
 2 & Roma      & Roma      & & $z_{2}$  & Cagliari  & 17 & 18 & 1\\
 3 & Napoli    & Lazio     & & $z_{3}$  & Cesena    & 19 & 19 & 0 \\
 4 & Lazio     & Napoli    & & $z_{4}$  & Chievo    & 18 & 16 & 2 \\
 5 & Sampdoria & Genoa     & & $z_{5}$  & Empoli    & 16 & 13 & 3 \\
 6 & Fiorentina& Milan     & & $z_{6}$  & Fiorentina& 6  & 8  & 2\\
 7 & Genoa     & Sampdoria & & $z_{7}$  & Genoa     & 7  & 5  & 2 \\
 8 & Palermo   & Fiorentina& & $z_{8}$  & Hellas    & 14 & 12 & 2\\
 9 & Udinese   & Inter     & & $z_{9}$  & Inter     & 11 & 9  & 2\\
10 & Milan     & Udinese   & & $z_{10}$ & Juventus  & 1  & 1  & 0\\
11 & Inter     & Torino    & & $z_{11}$ & Lazio     & 4  & 3  & 1\\
12 & Sassuolo  & Hellas    & & $z_{12}$ & Milan     & 10 & 6  & 4\\
13 & Torino    & Empoli    & & $z_{13}$ & Napoli    & 3  & 4  & 1\\
14 & Hellas    & Atalanta  & & $z_{14}$ & Palermo   & 8  & 15 & 7 \\
15 & Atalanta  & Palermo   & & $z_{15}$ & Parma     & 20 & 20 & 0\\
16 & Empoli    & Chievo    & & $z_{16}$ & Roma      & 2  & 2  & 0\\
17 & Cagliari  & Sassuolo  & & $z_{17}$ & Sampdoria & 5  & 7  & 2\\
18 & Chievo    & Cagliari  & & $z_{18}$ & Sassuolo  & 12 & 17 & 5\\
19 & Cesena    & Cesena    & & $z_{19}$ & Torino    & 13 & 11 & 2\\
20 & Parma     & Parma     & & $z_{20}$ & Udinese   & 9  & 10 & 1\\
\multicolumn{9}{c}{}\\
\cline{9-9}\multicolumn{9}{c}{}\\
\multicolumn{9}{r}{$D(A,P) = \sum_{i=1}^{20} |\tau_A(i)-\tau_P(i)|=38$} \\
\end{tabular}
\label{tab:rankings}
\end{center}
\end{table}

\paragraph{(b)} In the second case study, we are comparing the actual A and the predicted P final standings of the South American qualifiers for the 2010 FIFA World Cup, whose competing teams are listed in Tab.~\ref{tab:southam}. Following the same approach of case (a), we build the analogous Tab.~\ref{tab:rankings2}. Here the absolute total displacement is $D(A,P)=14$, apparently much smaller than in case (a), but the normalized distance $d(A,P)$ results $\frac{14}{\frac{10^2}{2}}=0.28$, showing instead a worse performance of the predictive algorithm $\mathcal{P}$ in case (b) compared to case (a).

\begin{table}[!b]
\begin{center}
\caption{{The set $T$ of teams playing in the South American qualifiers for the 2010 FIFA World Cup, alphabetically sorted.}}
\begin{tabular}{r|c||r|c}
Index & Team name & Index & Team name \\
\hline
$t_{1}$  & Argentina& $t_{6}$ & Ecuador\\  
$t_{2}$  & Bolivia& $t_{7}$ & Paraguay\\   
$t_{3}$  & Brazil& $t_{8}$ & Peru\\   
$t_{4}$  & Chile& $t_{9}$ & Uruguay\\   
$t_{5}$  & Colombia& $t_{10}$ & Venezuela\\   
\end{tabular}
\label{tab:southam}
\end{center}
\end{table}

\begin{table}[!t]
\begin{center}
\caption{{Actual (A) and predicted (P) table of Serie A 2014/15 after matchday 20, with the corresponding permutations $\tau_A$ and $\tau_P$ computed with respect to the set of teams $\mathcal{T}$.} In the last column the absolute displacement $|\tau_A-\tau_P|$ between A and P is reported for the corresponding team $z_i$, and its total is indicated in the last row.}
\begin{tabular}{r|c|cp{1cm}r|c|r|r|r}
Pos. & A & P & & T & Team & $\tau_A$ & $\tau_P$ & $|\tau_A-\tau_P|$ \\
\cline{1-3}\cline{5-9}
 1 & Brazil   & Argentina & & $t_{1}$  & Argentina & 4 & 1 & 3 \\
 2 & Chile    & Brazil    & & $t_{2}$  & Bolivia   & 9 & 9 & 0\\
 3 & Paraguay & Uruguay   & & $t_{3}$  & Brazil    & 1 & 2 & 1 \\
 4 & Argentina& Chile     & & $t_{4}$  & Chile     & 2 & 4 & 2 \\
 5 & Uruguay  & Colombia  & & $t_{5}$  & Colombia  & 7 & 5 & 2 \\
 6 & Ecuador  & Paraguay  & & $t_{6}$  & Ecuador   & 6 & 7 & 1\\
 7 & Colombia & Ecuador   & & $t_{7}$  & Paraguay  & 3 & 6 & 3 \\
 8 & Venezuela& Venezuela & & $t_{8}$  & Peru      & 10& 10& 0\\
 9 & Bolivia  & Bolivia   & & $t_{9}$  & Uruguay   & 5 & 3 & 2\\
10 & Peru     & Peru      & & $t_{10}$ & Venezuela & 8 & 8 & 0\\
\multicolumn{9}{c}{}\\
\cline{9-9}\multicolumn{9}{c}{}\\
\multicolumn{9}{r}{$D(A,P) = \sum_{i=1}^{10} |\tau_A(i)-\tau_P(i)|=14$} \\
\end{tabular}
\label{tab:rankings2}
\end{center}
\end{table}


\section*{Results}
In what follows, we will estimate the total number of earned points by a team, by mean of a linear model trained on the first $n-t_s$ matches of the seasons, for several values of $t_s$, for $n$ the total number of matches in the season. 
Furthermore, we will derive, for each championship, the estimate final league table to be compared with the actual standing.

\subsection*{Team performance prediction}
For the 7768 seasonal time series $T$, we estimate $\bar{T}_n$ for $t_s=1,\ldots,20$, with a linear, quadratic and cubic model.
As a first result, the linear model performs significatively better than the quadratic and cubic models, regardless of the length of the test set $t_s$.
As an example, consider the difference $|\bar{T}_n-T_n|$ across all 7768 series: for the linear model, the average is 4.652 with confidence interval (4.634, 4.672), while the same figures for the quadratic and cubic models are, respectively, 8.966 (8.913, 9.014) and 27.760 (27.530, 28.011). 
A paired t-test on all 7768 series between each couple of linear/quadratic/cubic models validate the hypothesis 
$|\bar{T}_n-T_n|_\textrm{linear} \le  |\bar{T}_n-T_n|_\textrm{quadratic} \le |\bar{T}_n-T_n|_\textrm{cubic}$ with $p$-value less than $10^{-16}$.  
Thus, in what follows, we will only discuss linear models.  

As a comparison, a null model obtained by applying a linear regressor to $10^5$ randomly generated time series of match results yields $\overline{|\bar{T}_n-T_n|}=4.993$ with confidence interval (4.666, 5.303), indicating that, globally, the considered real sequences are only slightly more structured than random.
Thus, a linear model applied either to a true or a random result sequence for a team during a season is expected to predict the final amount of points with an error of less than 5 points.

\begin{figure}[!t]
\begin{center}
\includegraphics[width=0.9\textwidth]{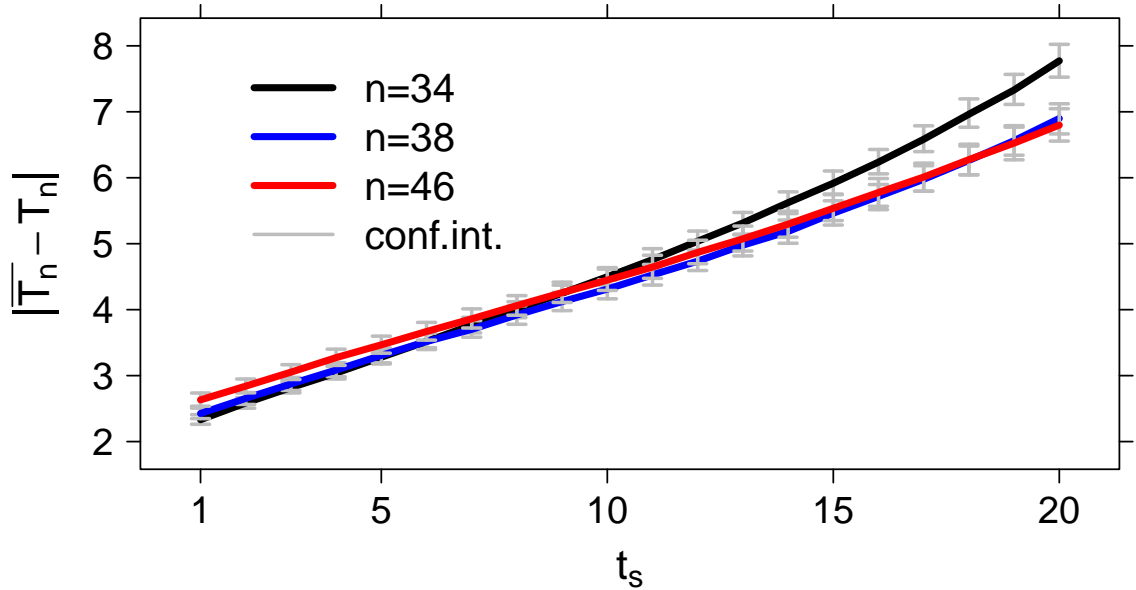}
\caption{{$\overline{|\bar{T}_n-T_n|}$} for increasing values of $t_s$ from 1 to 20 in the three cases $n=34, 38, 46$, with 95\% Student's bootstrap confidence intervals (in grey).}
\label{fig:n}
\end{center}
\end{figure}

We investigate now the value $\overline{|\bar{T}_n-T_n|}$ on a set of disaggregated covariates, including the total number of championship rounds, the length of the training or test portion, the country, the team, etc.
In Fig.~\ref{fig:n} we show the difference between the predicted and the real final amount of points $\overline{|\bar{T}_n-T_n|}$ for increasing values of $t_s$ from 1 to 20 in the three cases $n=34, 38, 46$ which collect most of the series (30\%, 25\% and 21\% respectively).
As evidenced by the graph, the difference between diverse values of $n$ is very small, and the overall performance of the linear models are quite good, even if a large portion of the results are excluded from the training set: \textit{e.g.}, for $t_s=10$, the average error is limited to 4.4 points for championships of any duration.

\begin{figure}[!t]
\begin{center}
\includegraphics[width=0.9\textwidth]{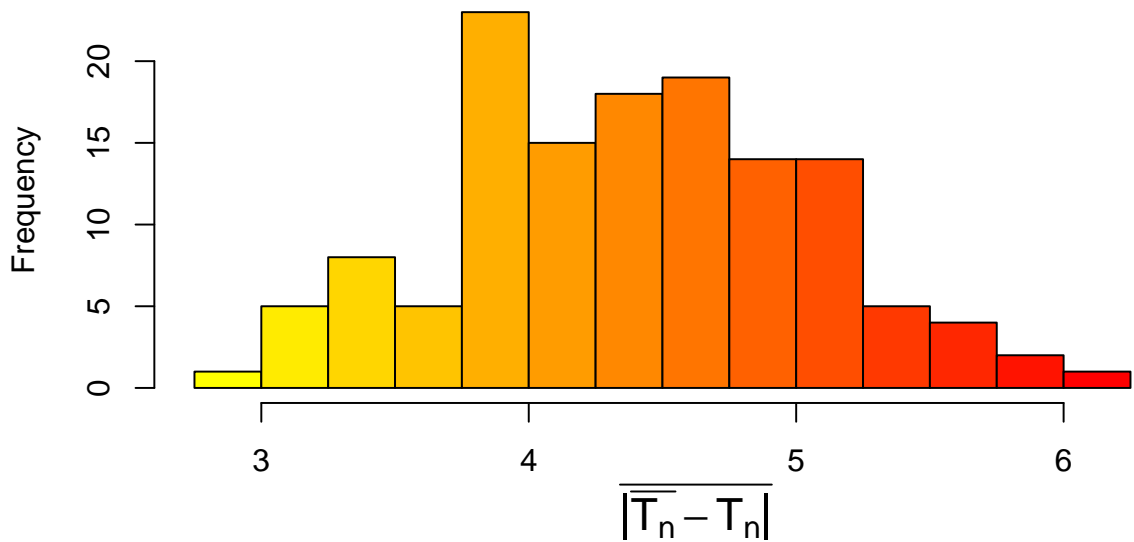}
\caption{{Histogram of $\overline{|\bar{T}_n-T_n|}$} for the set $S$ of 231 teams having more presences (18 or more seasons out of 21).}
\label{fig:histogram}
\end{center}
\end{figure}

\begin{figure}[!b]
\begin{center}
\centering\includegraphics[width=0.9\textwidth]{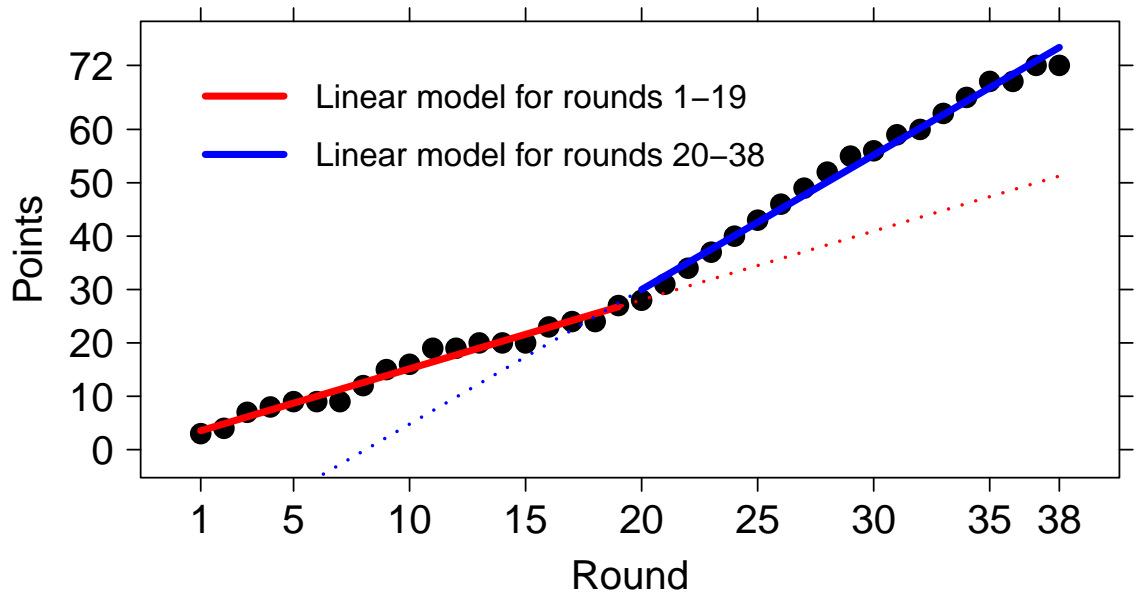}
\caption{{Points earned by FC Barcelona (black dots) in La Liga 2003/04} and the corresponding linear models for the first (red line) and the second (blue line) half of the season.}
\label{fig:barca}
\end{center}
\end{figure}

Consider now the (linear) predictivity ($\overline{|\bar{T}_n-T_n|}$ for $t_s=10$) of the set $S$ of 231 teams which are more present (18 or more seasons out of 21) in the available data: in Fig.~\ref{fig:histogram} the histogram is shown of the average differences between prediction and actual values. 
The set of values $\overline{|\bar{T}_n-T_n|}$ for $S$ is gaussian-like, with range $[2.75,6.19]$, (min and max corresponding to Sporting Braga and Oxford respectively) and mean and median $\approx 4.4$: smaller values indicate a more linear behavior of a team throughout all the considered seasons, while larger values mark the presence of one or more seasons where the sequence of results had a non-linear trend.
In Tab.~\ref{tab:top10} the values $\overline{|\bar{T}_n-T_n|}$ are listed for the top10 UEFA ranking teams (current standing at November 2015).
Among a number of teams such as Bayern, Chelsea and Juventus whose linear trend is quite consistent through all the considered seasons ($\overline{|\bar{T}_n-T_n|}<4)$, Barcelona's case emerges. 
Barcelona's high value $\overline{|\bar{T}_n-T_n|}=5.86$ is due to a number of seasons (1993/94, 2002/03, 2005/06, 2003/04, 2007/08, 2008/09, 2010/11) where the seasonal trend was markedly non-linear, mostly because the last matches followed a very different pattern from the initial part of the season.
As an example, consider the situation in the campaign 2003/04 as shown in Fig.~\ref{fig:barca}: the seasonal pattern is not linear, but it is piecewise linear, with the first and second half of the campaing following two distinct linear approximations whose corresponding slopes are respectively 1.29 and 2.52, thus in ratio almost 1:2. 

\begin{table}[!b]
\begin{center}
\caption{{$\overline{|\bar{T}_n-T_n|}$ for the top10 UEFA ranking teams at November 2015 for $t_s=10$.}}
\begin{tabular}{rlr|rlr}
1 & Real Madrid CF 		& 4.48	&6 & SL Benfica 			& 4.35 \\
2 & FC Bayern M{\"u}nchen 	& 3.86	&7 & Borussia Dortmund 			& 4.81 \\
3 & FC Barcelona 		& 5.86	&8 & Juventus 				& 3.71 \\
4 & Chelsea FC 			& 3.43	&9 & Paris Saint-Germain 		& 4.14 \\
5 & Club Atl{\'e}tico de Madrid & 5.09	&10 & Arsenal FC			& 4.57 \\
\end{tabular}
\label{tab:top10}
\end{center}
\end{table}

Furthermore, differences between various countries are small, for every value of $t_s$: as an example, for $t_s=10$, the values of $\overline{|\bar{T}_n-T_n|}$ ranges between 4.24 for Portugal to 4.69 for The Netherlands.
 
Finally, differences between teams ending in different zones of the final standing are also small: for $t_s=10$, the values (with confidence intervals) of $\overline{|\bar{T}_n-T_n|}$ for all teams finishing first to fifth is 4.32 (4.19, 4.46), for all teams filling the bottom 5 positions is 4.21 (4.07, 4.35), while for the teams in the 5 positions at the middle of the table the corresponding values are slightly larger 4.54 (4.39, 4.67) indicating a less precise linear predictivity for these teams.

\subsection*{Championship outcome prediction}
Let us now consider predicting the final outcome not of a single team, but rather of an entire championship.
As a performance measure, we use the normalized total absolute displacement $d$ outlined in Methods.

\begin{figure}[!t]
\begin{center}
\includegraphics[width=0.95\textwidth]{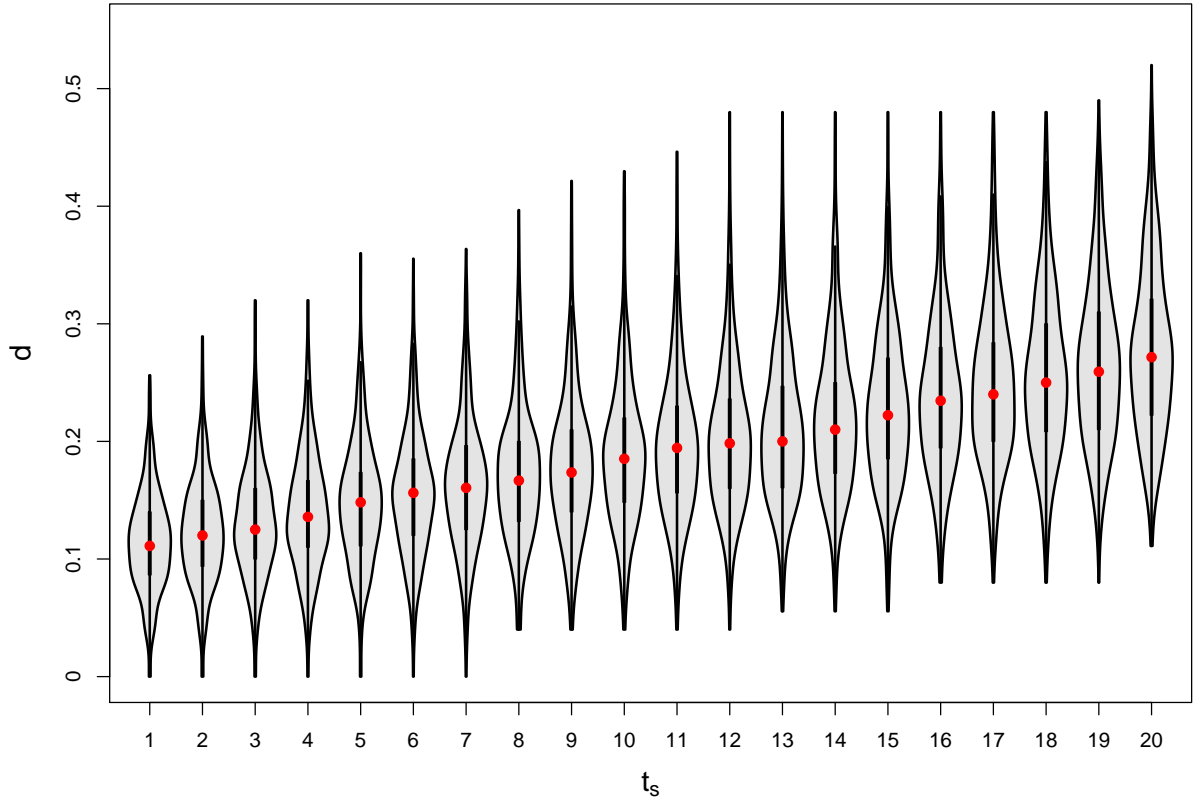}
\caption{{Violin plot of normalized total absolute displacement $\pmb{d}$ as a function of $\pmb{t_s}$ averaged over the 425 championships, with distribution (grey), median (red dots) and boxplot (inner black line).}}
\label{fig:totdisp}
\end{center}
\end{figure}

As a first result, in Fig.~\ref{fig:totdisp} we plot, for each $1\leq t_s\leq 20$,  the distribution of the normalized total absolute displacements $d$ for the 425 championships included in the considered dataset. 
The 95\% Student's bootstrap confidence intervals $[l,u]$ are not reported in the figure because they are too narrow: for each $t_s$, we have $[l,u]\subset [\frac{\bar{d}}{1.038},1.037\bar{d}]$.
As a function of $t_s$, the median of $d$ is very close to $\bar{d}$ (the ratio between the mean and the median of $d$ ranges between 0.988 and 1.057), and it has an almost linear trend significatively smaller than the null model value $\approx\frac{2}{3}$ even for large values of $t_s$.
For example, for $t_s=10$ we have $\bar{d}=0.1874$ which, for a tournament with 20 teams, means that, in average, the linear model can guess the final ranking of each team with an error of 1.874 positions.
In 25 cases (with $t_s\leq 7$), the actual final ranking was perfectly predicted by the linear model.

No significative difference in the table prediction performance is also detected when comparing the top leagues (Premier League, Serie A, Ligue 1, La Liga, Bundesliga, Eredivisie, Primeira Liga) with all the other considered leagues: $\bar{d}$ for the former championships is 0.184 (0.175, 0.192), while for the latter is 0.189 (0.180, 0.198).

A crucial task championship outcome prediction is to forecast the final top and bottom of the table, that is, the teams qualifying for European tournaments (UCL, EL) and the teams facing relegation.
Define the True Positive Rate (TPR) as the fraction of championships (out of 425) where all the teams finishing in top-k (or bottom-k) positions were correctly predicted by a linear model.
In Tab.~\ref{tab:topbottom} the TPR is shown for increasing $t_s=1,\ldots,20$, for the first/last $k=3$ and $k=6$ positions.
Overall, the performance of the linear model is quite good for a wide range of values of $t_s$: for $t_s<10$, the TPR is larger than 0.9 for all cases.
Moreover, predictions for $k=3$ is slighly noisier than $k=6$, while in both cases predicting the bottom of the table is slighlty harder than guessing the top teams.

\begin{table}[!t]
\begin{center}
\caption{{True Positive Rate of linear prediction of top/bottom-$k$ (Tk,Bk) teams for $k=3$ and $k=6$.}}
\begin{tabular}{r|rrrrrrrr}
$t_k$ & T3 & \% & B3 & \% & T6 & \% & B6 & \%\\
\hline
 1&422&0.992&414&0.974&425&1.000&422&0.993\\
 2&421&0.991&411&0.967&425&1.000&422&0.993\\
 3&420&0.988&410&0.965&425&1.000&422&0.993\\
 4&418&0.984&403&0.948&425&1.000&421&0.991\\
 5&417&0.981&402&0.946&425&1.000&420&0.988\\
 6&416&0.979&401&0.944&425&1.000&420&0.988\\
 7&413&0.972&396&0.932&425&1.000&420&0.988\\
 8&412&0.969&390&0.918&425&1.000&418&0.984\\
 9&409&0.962&383&0.901&425&1.000&415&0.976\\
10&405&0.953&379&0.892&425&1.000&413&0.972\\
11&403&0.948&373&0.878&425&1.000&411&0.967\\
12&401&0.944&374&0.880&424&0.998&411&0.967\\
13&396&0.932&370&0.871&424&0.998&409&0.962\\
14&396&0.932&363&0.854&424&0.998&405&0.953\\
15&392&0.922&356&0.838&424&0.998&406&0.955\\
16&384&0.904&351&0.826&423&0.995&406&0.955\\
17&383&0.901&347&0.816&422&0.993&403&0.948\\
18&375&0.882&345&0.812&419&0.986&401&0.944\\
19&368&0.866&338&0.795&418&0.984&402&0.946\\
20&363&0.854&332&0.781&415&0.976&398&0.936
\end{tabular}
\label{tab:topbottom}
\end{center}
\end{table}

\subsection*{Example: EPL 12/13} We conclude with a particularly favorable example (English Premier League 2012/13 relegation zone) where the linear model predictivity is better than the more complex combinations of algorithm and human knowledge which translate into the odds offered by betting services.
In Tab.~\ref{tab:odds} the corresponding relegation odds are reported for six betting agencies, namely (B1) \textit{Betting Expert}~\cite{moncrieff13premier}, (B2) \textit{bwin}~\cite{curran13premier}, (B3) \textit{Bet365}~\cite{jackson13who}, (B4) \textit{Ladbrokes}~\cite{james12premier}, (B5) \textit{SportBookReview}~\cite{richardson12soccer} and (B6) \textit{William Hill}~\cite{riley12premier}, together with the average odds.
Although the betting odds were suggesting for instance Norwich and Southampton as likely candidates (with 2.50 and 2.16 average odds), quite unexpectedly (average odds 4.95) Queen's Park Rangers suffered relegation instead.
In this case, the linear model performs effectively, consistently predicting QPR, Reading and Wigan as the relegated teams, for each $t_s=1,\ldots,20$.

\begin{table}[!b]
\begin{center}
\caption{{Relegation odds for 6 betting agencies for English Premier League 2012/13.} Last column shows the averae odds. In boldface the three relegated teams. (B1) {Betting Expert}~\cite{moncrieff13premier}, (B2) {bwin}~\cite{curran13premier}, (B3) {Bet365}~\cite{jackson13who}, (B4) Ladbrokes~\cite{james12premier}, (B5) SportBookReview~\cite{richardson12soccer} and (B6) William Hill~\cite{riley12premier}}
\begin{tabular}{l|rrrrrr|r}
Team & B1 & B2 & B3 & B4 & B5 & B6 & Mean \\
\hline
Norwich 	&2.60         & 1.75        & 1.50        & 1.50        & 6.00        & 1.63 	        & 2.50 \\
{\bf QPR}	&{\bf 7.20}   & {\bf 4.50}  &{\bf 5.00}   &{\bf 4.00}   &{\bf 4.00}   & {\bf 5.00}	& {\bf 4.95} \\
{\bf Reading} 	&{\bf 2.70}   & {\bf 1.00}  &{\bf 1.10}   &{\bf 1.10}   &{\bf 4.00}   & {\bf 1.10}	& {\bf 1.83} \\
Southampton 	&2.40         & 1.20        & 1.38        & 1.25        & 5.50        & 1.25		& 2.16 \\
Swansea 	&3.10         & 2.00        & 2.25        & 2.00        & 9.00        & 1.75		& 3.35 \\ 
West Bromwich   &4.40         & 3.50        & 3.50        & 3.33        & 3.33        & 4.50		& 3.76 \\
West Ham 	&4.00         & 2.20        & 2.00        & 2.25        & 10.00       & 1.63	        & 3.68 \\
{\bf Wigan} 	&{\bf 2.80}   &{\bf 1.75}   &{\bf 1.5}    &{\bf 1.63}   &{\bf 6.00}   & {\bf 1.64}	& {\bf 2.55} 
\end{tabular} 
\label{tab:odds}
\end{center}
\end{table}

\section*{Conclusions}
An high level of linearity may come unexpected when dealing with football results, where a large number of confounding factors concur in influencing the outcome of both a single match and an entire tournament.
Here we show that, when considering long tournaments like national championships, linear trends are quite widespread, and linear models can also work as effective predictors.
In particular, we tested the linear forecast of the total number of earned points by a team during a season, and the final team ranking in the table, where the model is trained only on the initial portion of the season.
In both cases, we demonstrated that even such a minimalist approach and without using historical data can achieve good predictive performances.

\bibliography{jurman15seasonal}
\end{document}